\documentclass[%
 reprint,
superscriptaddress,
onecolumn,
 amsmath,amssymb,
 aps,
]{revtex4-2}

\usepackage{amsfonts}
\usepackage{amssymb}
\usepackage{amsmath}
\usepackage{calc}
\usepackage{subfigure}
\usepackage{graphicx}
\usepackage{epstopdf}
\usepackage{dcolumn}
\usepackage{bm}
\usepackage{color}
\usepackage{txfonts}
\usepackage[dvipsnames]{xcolor}
\usepackage[colorlinks,citecolor=blue]{hyperref}

\begin{document}

\title{Fast transport of Bose-Einstein condensates in anharmonic traps}

\author{Jing Li}
\email{jli@ucc.ie}
 \affiliation{Department of Physics, University College Cork, Cork, Ireland, T12 H6T1}

\author{Xi Chen}
\affiliation{Department of Physical Chemistry, University of the Basque Country UPV/EHU, Apartado 644, 48080 Bilbao, Spain}
\affiliation{EHU Quantum Center, University of the Basque Country UPV/EHU, 48940 Leioa, Spain}

\author{Andreas Ruschhaupt}
 \affiliation{Department of Physics, University College Cork, Cork, Ireland, T12 H6T1}

\date{\today }

\begin{abstract}
We present a method to transport Bose-Einstein condensates (BECs) in anharmonic traps and in the presence of atom-atom interactions in short times without residual excitation. Using a combination of a variational approach and inverse engineering methods, we derive a set of Ermakov-like equations that take into account the coupling between the
center of mass motion and the breathing mode. By an appropriate inverse engineering strategy of those equations, we then design the trap trajectory to achieve the desired boundary conditions. Numerical examples for cubic or quartic anharmonicities are provided for fast and high-fidelity transport of BECs. Potential applications are atom interferometry and quantum information processing.
\end{abstract}

\maketitle


\section{Introduction}

The accurate manipulation of ultracold atoms is a key pre-requisite to implement 
quantum technologies within atomic, molecular and optical
science \cite{milburn1996quantum}.
In particular, the transport of individual atoms and of thermal or Bose-condensed clouds using moving traps has been demonstrated in many experiments  \cite{Nature,Esslinger,Ketterle,Meschede,Lahaye,David08,Calarco09,James,Bowler,Walther,Donner} for different goals in quantum information processing and metrology.
In all quantum technologies, preserving quantum coherence and achieving high final fidelities
in short times is of crucial importance. 
One possibility is called shortcuts to adiabaticity (STA) \cite{TORRONTEGUI2013117,STAreview}
which provides a toolbox to control both the internal and external degrees of freedom of a quantum system in {faster-than} adiabatic 
times.

Various shortcuts to adiabatic transport have been proposed: Lewis-Riesenfeld invariant-based inverse engineering \cite{Torrontegui2012,Erik11,Jing17,Qi2015,transOCT12}, enhanced STA scheme \cite{chrisPRR,chris2021,Whitty_2022}, the Fourier optimization \cite{Uli14}, 
fast-forward scaling method \cite{Masuda10,Masuda12}, and the counter-diabatic driving \cite{Adol14}
have been theoretically put forward, and experimentally demonstrated for various systems \cite{Adol16,David08,Bowler,Walther}. 
The possibility to operate with short times not only reduces the sensitivity to low frequency noise, but also allows for improved measurement statistics in the total time available for the experiment.

Different approaches for transporting particles have been implemented. 
Neutral atoms have been transported as 
Bose-Einstein condensates (BECs) \cite{Nature}, thermal atomic clouds \cite{Hommelhoff}, or individually \cite{Meschede}, using magnetic
or optical traps. 
The commonly used traps for ultracold atoms based on electromagnetic fields are never perfectly harmonic.
{The weak cubic anharmonicity plays a role when a BEC is transported perpendicular to the atom chip surface \cite{Corgier2018}.}
The quartic anharmonicity is significant when approximating the potential of an optical tweezers for transport \cite{David08,Erik11}.
Thus cancelling the anharmonic contributions of the trapping potential is vital for useful control schemes and is already a difficult technical challenge for a static trap \cite{Lobser2015}. 


Anharmonicities can have an important impact on the dynamics as observed in  atom cooling \cite{Rabitz}, collective modes \cite{Liu}, or wave packet dynamics \cite{Lenoeuf}. In most cases, the anharmonic traps are
considered as a perturbation of a harmonic one.
Perturbation theory has been used to  design  shortcut protocols
for expansion/compression \cite{Lu14PRA} and transport \cite{Mikel13}. 
Of course the results are limited by the premises of perturbation theory, i.e., by small anharmonicities.  
Considering a non-perturbative scenario is thus of much interest.

In this paper, we propose to inverse engineer rapid and robust transport of an interacting Bose-Einstein condensate (BEC) in  anharmonic traps using a variational approach. The method relies on a variational formulation of the dynamics to derive a set of coupled Ermakov-like and Newton-like equations, from which
the trap trajectory is inferred interpolating between the desired boundary conditions. In Sec.~\ref{sec1}, we explain the variational formalism. In Sec.~\ref{sec2}, we work out the explicit solutions for quartic and cubic anharmonicities of the confining potential, and  illustrate the efficiency of the method with various numerical examples. In section \ref{sec3}, we will discuss the results.

\section{Model, Hamiltonian and Method}
\label{sec1}

\begin{figure}[]
\begin{center}
{\includegraphics[width=0.35\textwidth,height=4cm]{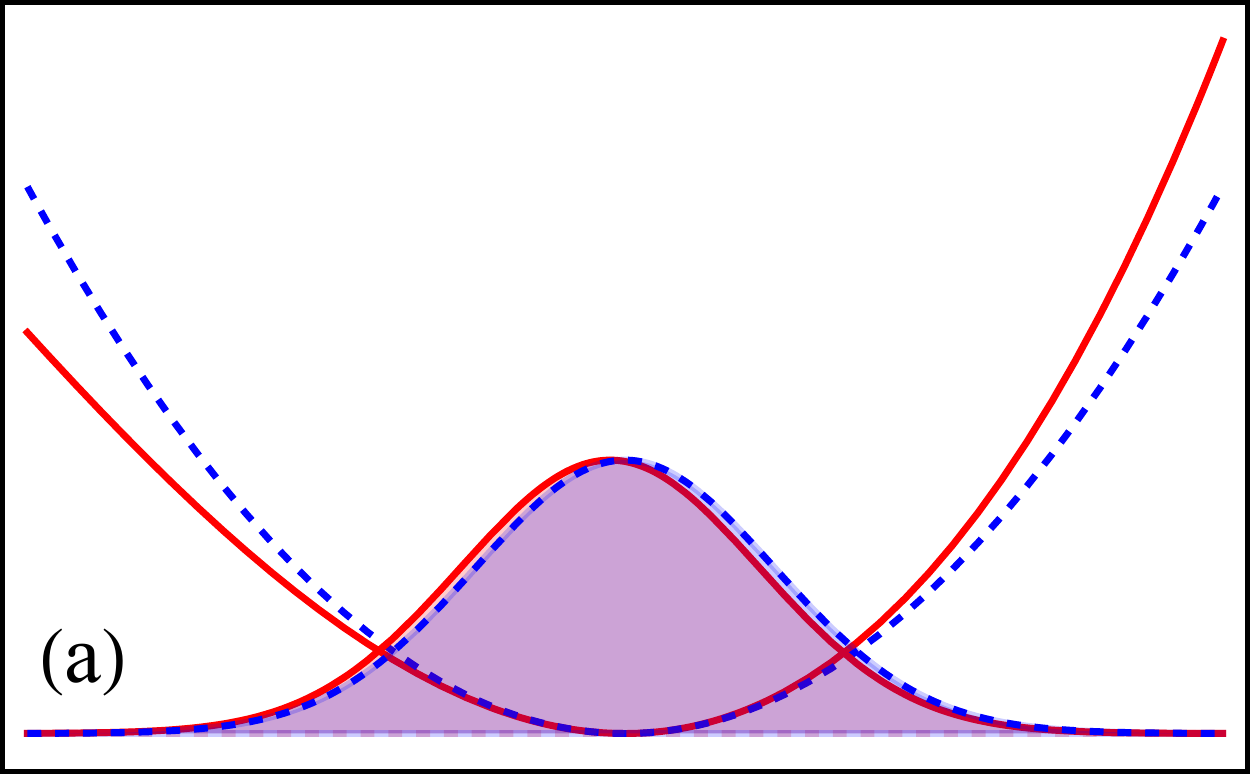}}
{\includegraphics[width=0.35\textwidth,height=4cm]{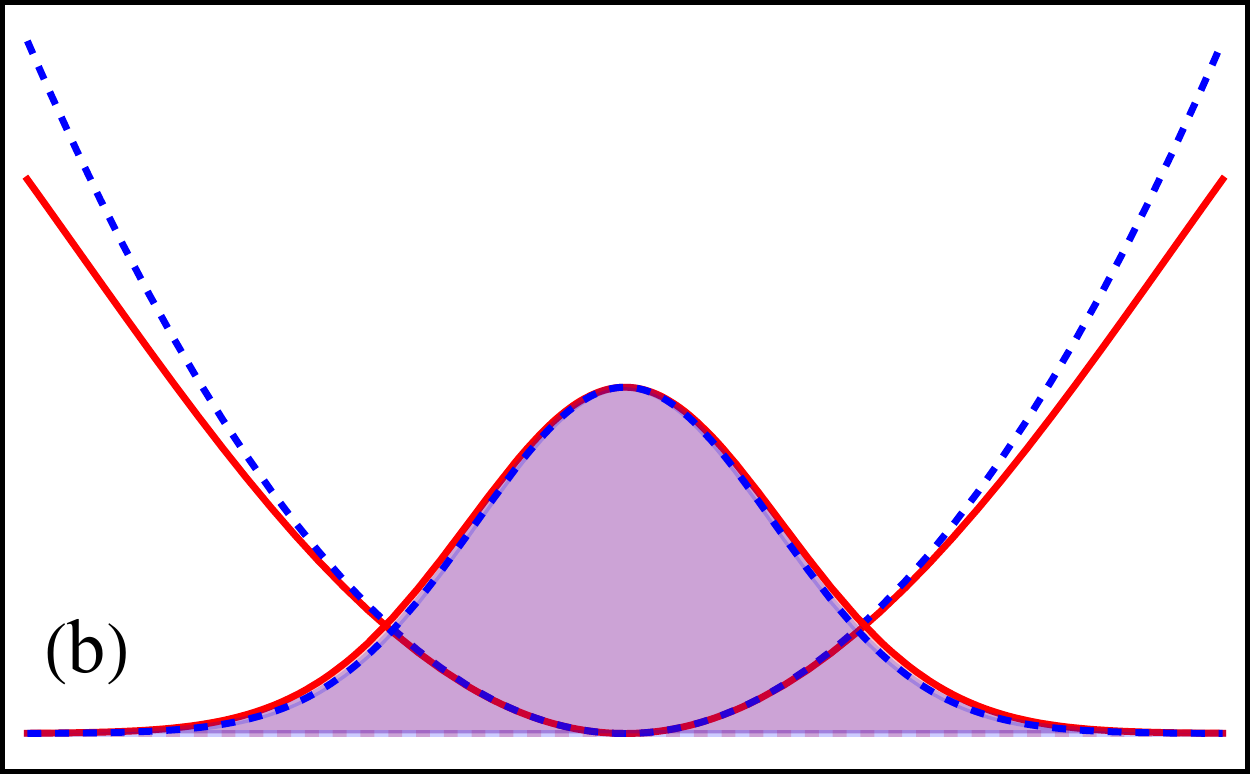}}
\caption{Cubic (a) and quartic (b) anharmonic potentials (red solid lines) compared to the harmonic counterparts 
(blue dashed lines). The ground states are plotted for different potentials.}
\label{fig1-setting}
\end{center}
\end{figure}

For a cigar-shaped trap with strong transverse confinement, e.g. $\omega_{\bot}>>\omega$, it is appropriate to consider a 1D dimensionless formula by freezing the transverse dynamics to the respective ground state and integrating over the transverse variables\cite{PhysRevA.70.023604}.
{The effective atomic interaction is denoted by $g=2 a_s \omega_{\bot} N/\omega a_{ho}$, with $a_s$ the interatomic scattering length and $a_{ho}=\sqrt{\hbar/(m\omega)}$.}
The resulting dimensionless form of Gross-Pitaevskii equation (GPE) \cite{cctdgo} can be written as
\begin{equation}
i \frac{\partial \psi(x,t)}{\partial t}= \left[-\frac{1}{2} \frac{ \partial^2}{\partial x^2}+V(x,t){+}g\vert \psi(x,t) \vert^2 \right] \psi(x,t),
\label{GPE}
\end{equation}
where
\begin{equation}
V(x,t)= \frac{1}{2} \left[x-x_0(t)\right]^2+\frac{\kappa}{3!} [x-x_0(t)]^3+\frac{\lambda}{4!}[x-x_0(t)]^4,
\label{potential}
\end{equation}
where $\psi(x,t)$ is the axial wavefunction of the condenstate with normalization condition $\int^{+\infty}_{-\infty} |\psi(x,t)|^2 dx=N$.
The attractive and repulsive interactions are denoted by $g<0$ and $g>0$, respectively. 
The axial harmonic trap frequency is $\omega$. The potential center $x_0(t)$ is {time-dependent} for transport.
Note that the potential in Eq. (\ref{potential}) consists two types of anharmonicities,  one is cubic ($\kappa\ge 0$) and the other is quartic ($\lambda \ge 0$) anharmonicity, which is shown in Fig. \ref{fig1-setting}.

To apply the variational approach we first define an ansatz for the wave function with a few free parameters and evaluate the Lagrangian density. The minimization of the total Lagrangian with respect to the free parameters provides equations of motion for the free parameters \cite{Zoller}. This approach is equivalent to a moment method \cite{impens}.

We assume a general Gaussian ansatz,
\begin{eqnarray}
\label{anstz}
\psi(x,t)=A(t) \exp\left[-\frac{(x-x_c(t))^2}{2a(t)^{2}}\right] \exp\left[ib(t)(x-x_c(t))^2+ic(t)(x-x_c(t))+i\phi(t)\right]
\end{eqnarray}
where the time-dependent parameters $A(t)$, $a(t)$, $b(t)$, $c(t)$, and $\phi(t)$ represent respectively the amplitude, width, chirp,
velocity, and global phase. The wave function center of mass is $x_c(t)$. 
In the following we omit $t$ in those variables for simplification. The normalization condition yields $A=\sqrt{N/(a\sqrt{\pi})}$.

The Lagrangian density which corresponds to Eq. (\ref{GPE}) reads \cite{Zoller}
\begin{eqnarray}
\label{Lagdensity}
L=\frac{i}{2}\left(\frac{\partial \psi}{\partial t}\psi^{*}-\frac{\partial
\psi^*}{\partial t}\psi\right)-\frac{1}{2}\left|\frac{\partial \psi}{\partial x}%
\right|^2  {-}\frac{g}{2}|\psi|^4-V(x)|\psi|^2.
\end{eqnarray}
Inserting the ansatz (\ref{anstz}) into Eq.~(\ref{Lagdensity}), we find an effective Lagrangian \cite{Zoller} by
integrating the Lagrangian density over the whole coordinate space, $%
\mathcal{L}=\int^{+\infty}_{-\infty} L dx$. The Euler-Lagrange minimization is performed over $\mathcal{L}$ and with respect to the free parameters and the conditions $\delta \mathcal{L}/\delta \xi=0$ where $\xi=a, b, c$ or $x_c$.
Four coupled equations result for $(\dot{a}, \dot{b}, \dot{x}_c, \dot{c})$, are given by

\begin{eqnarray}
\label{adot}
&&\dot{a}=2ab, 
\\
\label{bdot}
&&\dot{b}=\frac{1}{2a^4}-\frac{1}{2}\left[1-4\kappa (x_0-x_c)+18\lambda(x_0-x_c)^2 \right]-2b^2{+}\frac{gN}{2\sqrt{2\pi}a^3},
\\
\label{cdot}
&&\dot{c}=(1+18\lambda a^2)(x_0-x_c)-2\kappa (x_0-x_c)^2+12\lambda (x_0-x_c)^3-\kappa a^2,
\\
\label{xcdot}
&&\dot{x_c}=c,
\end{eqnarray}
which can be condensed into two second order 
coupled equations for the width $a$ and the wavepacket center $x_c$,
\begin{eqnarray}
\label{engeering1}
&&\ddot{a}=\frac{1}{a^3}-a\left[1-4\kappa q+18\lambda q^2 \right]{+}\frac{g N}{\sqrt{2\pi}a^2}-9\lambda a^3,
\\
\label{engeering2}
&&\ddot{x}_c=(1+18\lambda a^2)q-2\kappa q^2+12\lambda q^3-\kappa a^2,
\end{eqnarray}
where $q=x_0-x_c$ is the displacement between the center of the harmonic term and the wavepacket. 
In Eq. (\ref{engeering1}) the center of mass motion $x_c$ is strongly coupled with the width $a$ of the wave function through the anharmonic terms of the confining potential.
When we consider an adiabatic transport such as $q=0$, one can see that the cubic anharmonicity $\kappa$ is strongly coupled with the width $a$ in Eq. (\ref{engeering2}). 
Alternatively, the quartic anharmonicity $\lambda$ will create breathing mode due to the strong coupling with the width $a$ in the case of $q=0$. 
The non-linearities introduced by atom-atom interactions do not generate any coupling with anharmonicities, as is known for harmonic traps \cite{Erik11}.

This system generalizes the 
structure found for harmonic traps via invariant-based inverse engineering \cite{Erik11}.
In the absence of anharmonicities ($\kappa =0$ and $\lambda=0$), the two  coupled equations Eqs. (\ref{engeering1}) and (\ref{engeering2}) reduce to
an Ermakov equation \cite{PRL104} and a Newton equation \cite{Erik11} for a single atom (or ion) or a BEC.
By contrast, Eq.~(\ref{engeering2}) for the trajectory of the center of mass $x_c$
can be  generically recovered from the Ehrenfest theorem, and is therefore immune to the precise shape of the ansatz.
In what follows we shall exploit these coupled equations
to inverse engineer
shortcut to adiabatic transport of BECs.

\section{Inverse engineering}
\label{sec2}
In this section, we focus on 
the fast and high fidelity transport of a BEC
from a stationary state at initial position $x_0(0)=0$ to a target state with $x_0(t_f)=d$ in a finite time $t_f$. The desired distance of potential is $d$.
We will consider the cases of cubic (see Sec. \ref{sec2}\ref{sec:cubic}) and quartic (see Sec. \ref{sec2}\ref{sec:quartic}) anharmonicities individually. 
In particular, the trajectory $x_0(t)$ of the potential center can be designed by using inverse engineering methods applied to the set of equations (\ref{engeering1}) and (\ref{engeering2}). 
Furthermore, we will provide numerical examples that confirm the effectiveness of the method.
\subsection{Cubic anharmonicity}
\label{sec:cubic}
%
%
%
\begin{figure}[t]
\begin{center}
\includegraphics[width=0.35\textwidth,height=4cm]{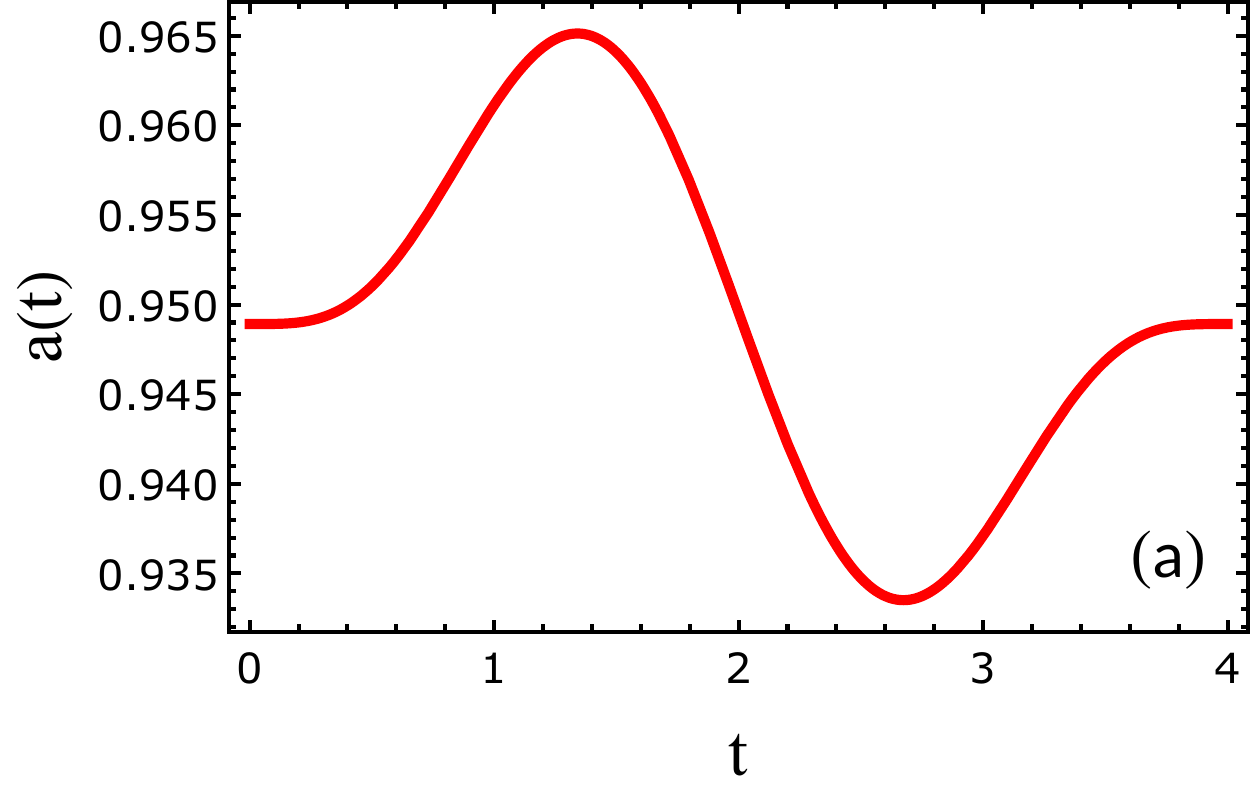}\quad
\includegraphics[width=0.35\textwidth,height=4cm]{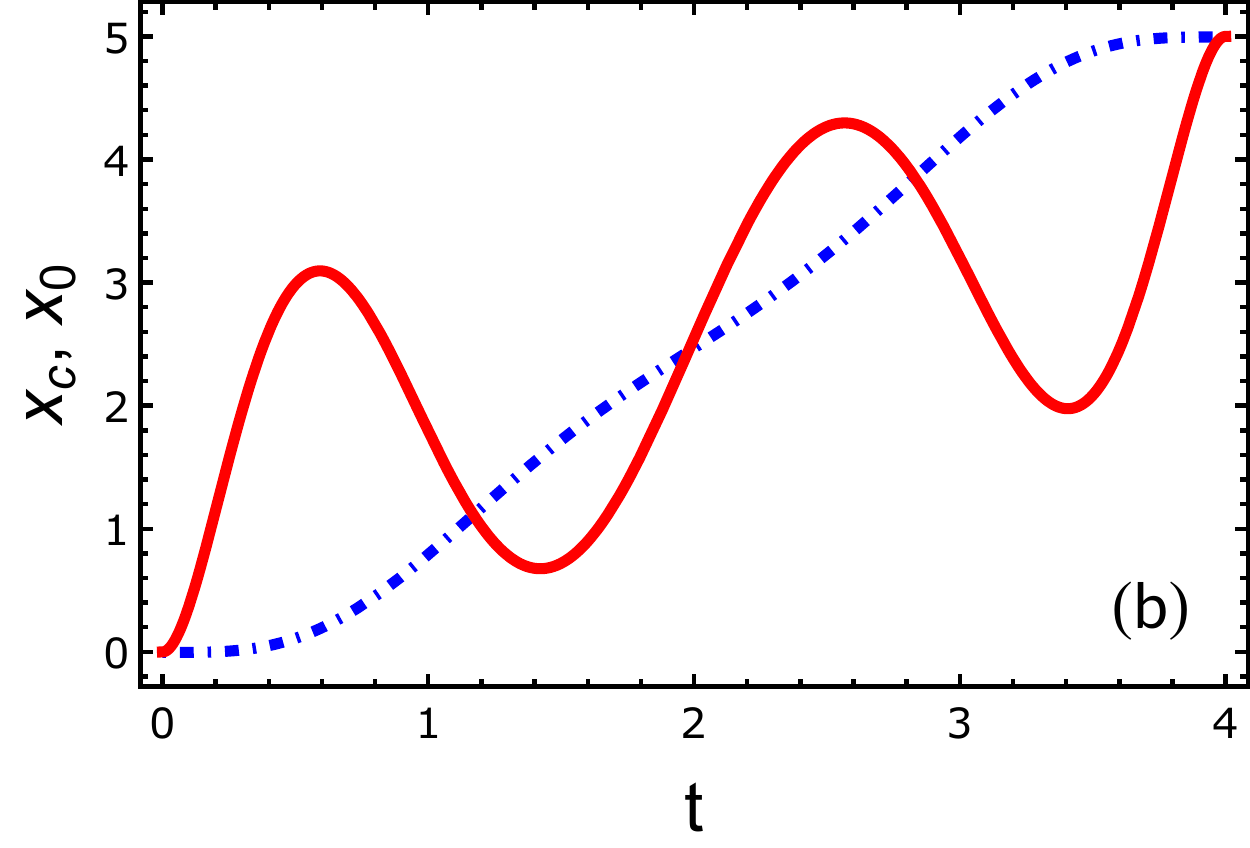}\\
\vspace*{-4mm}
(c) \includegraphics[width=0.6\textwidth]{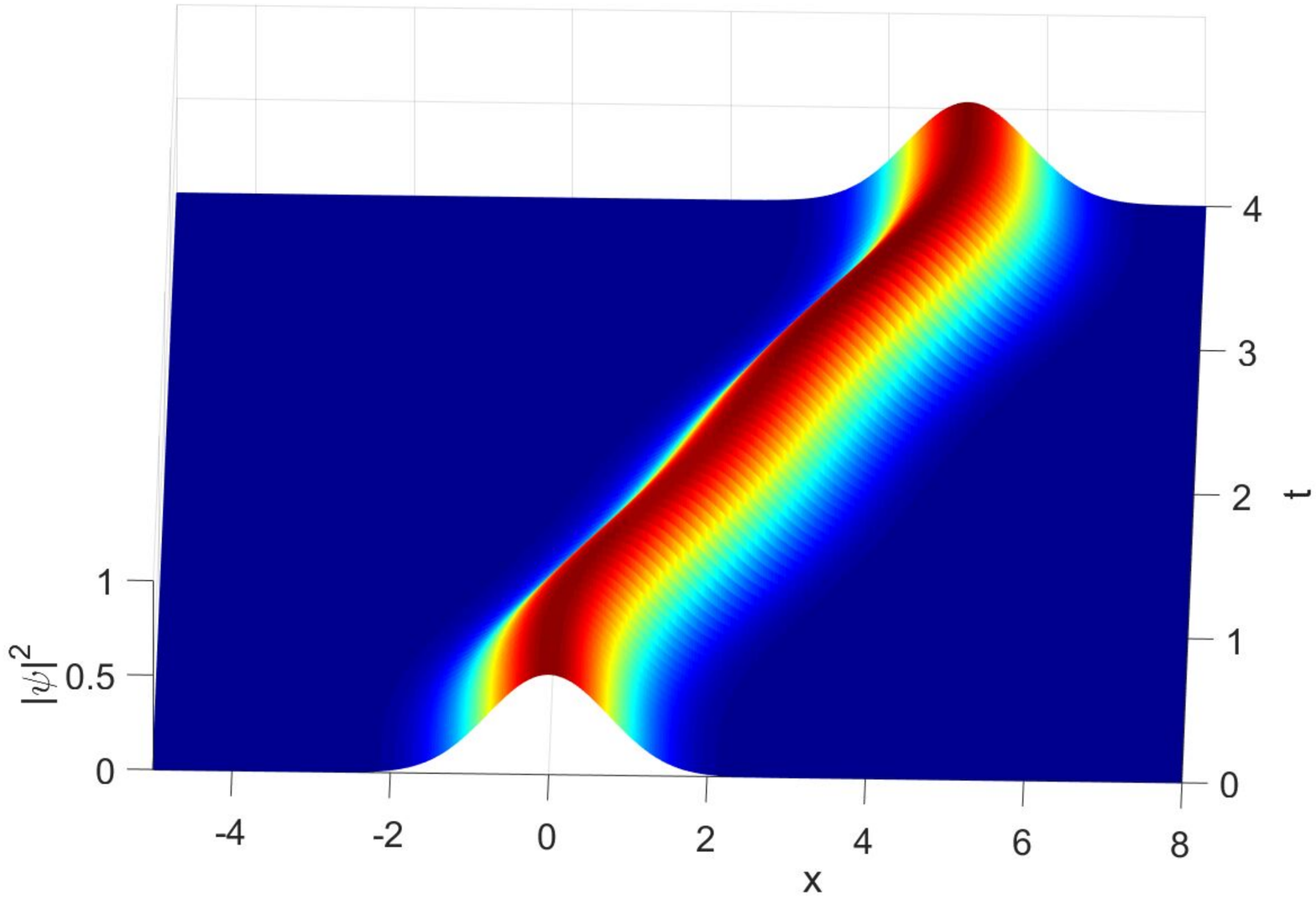}
\caption{Cubic anharmonicity: (a) width $a(t)$ with respect to time. 
(b) designed trajectories for trap center $x_0$ (red solid) and center of mass $x_c$ (dot-dashed blue).
The rest of parameters are $Q=4.5\times 10^{-3}$, $a_{0}=0.95$, $g=0.5$, $\kappa=0.02$,  $t_f=4$, and the distance $d=5$ of transport. (c) The corresponding time evolution  $|\psi_{STA}(x,t)|^2$. }
\label{figx3}
\end{center}
\end{figure}

Let us consider a potential with cubic anharmonicity \cite{Qi2015},
\begin{equation}
  V(x,t)=\frac{1}{2}(x-x_0)^2+\frac{1}{3!}\kappa (x-x_0)^3.  
\end{equation}

When $\kappa \neq 0$ and $\lambda =0$, we substitute the condition $\ddot{x}_c=\ddot{x}_0-\ddot{q}$ into the coupled differential equations (\ref{engeering1}) and (\ref{engeering2}), which can be simplified into
\begin{eqnarray}
\label{x3a}
&&\ddot{a}=\frac{1}{a^3}-a{+}\frac{gN}{\sqrt{2\pi}a^2}+4\kappa a q,
\\
&&\ddot{q}=\ddot{x}_0-q+\kappa a^2+2\kappa q^2.\label{x3x}
\end{eqnarray}

The second equation may be regarded as a second order differential
equation for $q$. 
We require that both initial and final states are stationary states.
First we can calculate the initial and final conditions for the function $q$ which are $q(0)=q(t_f)=Q$.  By imposing $\ddot{x}_0-\ddot{q}=0$ in Eq. (\ref{x3x}), one obtains 
\begin{equation}
\label{eq:cubicQ}
Q=\frac{1-\sqrt{1-8a_{0}^2\kappa^2}}{4\kappa},
\end{equation}
where $a_{0}$ denotes the initial and final widths, which are equal. 
Note that the difference $Q$ is caused by the asymmetricity of the cubic anharmonic potential. 
Substituting Eq. (\ref{eq:cubicQ}) into Eq. (\ref{x3a}), we can obtain the initial width $a_0$ as well as the final width by imposing $\ddot{a}=0$,
\begin{equation}
 \label{widtha0x3}
\frac{1}{a_{0}^3}-a_{0}{+}\frac{g N}{\sqrt{2\pi}a_{0}^2}+2\kappa a_{0}Q=0.  
\end{equation}
The value of $a_0$ is numerically obtained by solving Eq. (\ref{widtha0x3}), which is dependent on the values of the nonlinearity $g$ and anharmonicity strength $\kappa$. The width $a_0$ increases when the system has either repulsive interaction or cubic anharmonicity.

Now we use inverse engineering according to the following steps.
We may recall that the initial and final states are stationary states with width $a_0$
without excitations at the final time.
Then we can set up the boundary conditions for width $a$ according to Eq. (\ref{x3a}):
\begin{eqnarray}
\label{BCsx3}
&&a(0)=a_0,\quad a(t_f)=a_0, \\
&&\ddot{a}(0)=0, \quad \ddot{a}(t_f)=0.
\end{eqnarray}
Since that the chirp and velocity terms satisfy $b(0)=b(t_f)=0$ and $c(0)=c(t_f)=0$ in Eqs. (\ref{adot}) and (\ref{xcdot}), respectively, one can find the conditions from Eqs. (\ref{bdot}) and (\ref{cdot}) that
\begin{eqnarray}
\dot{a}(0)=0,\quad \dot{a}(t_f)=0 \\
\dot{q}(0)=0, \quad \dot{q}(t_f)=0,
\end{eqnarray}
In addition, the boundary conditions for $x_0$ in Eq. (\ref{x3x}) are imposed by
\begin{equation}
\label{bcx0}
x_0(t_f)=d,\;\dot{x}_0(t_f)=0.
\end{equation}
Then we set a ninth-order polynomial for $a(t)=\sum_{n=0}^9 a_n t^n$ and fix the parameters by satisfying all the boundary conditions of Eqs. (\ref{BCsx3})-(\ref{bcx0}).
An example of the designed function $a$ is shown in Fig. \ref{figx3} (a). Once we obtain the function $a$, one can easily get the function $q$ in Eq. (\ref{x3a}).
Finally, $x_0$ and $x_c$ can be expressed easily in terms of the width $a$ and $q$ which is shown in Fig. \ref{figx3} (b).  
Note that we fix values of $g$, $\kappa$ and final time $t_f$ in the example. Fig. \ref{figx3} (a) shows the wavepacket undergoes a slight breathing and finally returns to the initial width during the non-adiabatic process. {This breathing phenomena is due to the coupling term between anharmonicity $\kappa$ and width $a$ in Eq. (\ref{x3a}):
with $\kappa=0$, the solution of Eq. (\ref{x3a}) will be a constant width $a$. }
Fig. \ref{figx3} (b) illustrates that the trap trajectory oscillates from the initial position and then returns to the desired position at $x_0=d$. 
{The corresponding time-evolution $|\psi_{STA}(x,t)|^2$ is shown in Fig. \ref{figx3}(c).}

\begin{figure}[t]
\begin{center}
\scalebox{0.65}[0.65]{\includegraphics{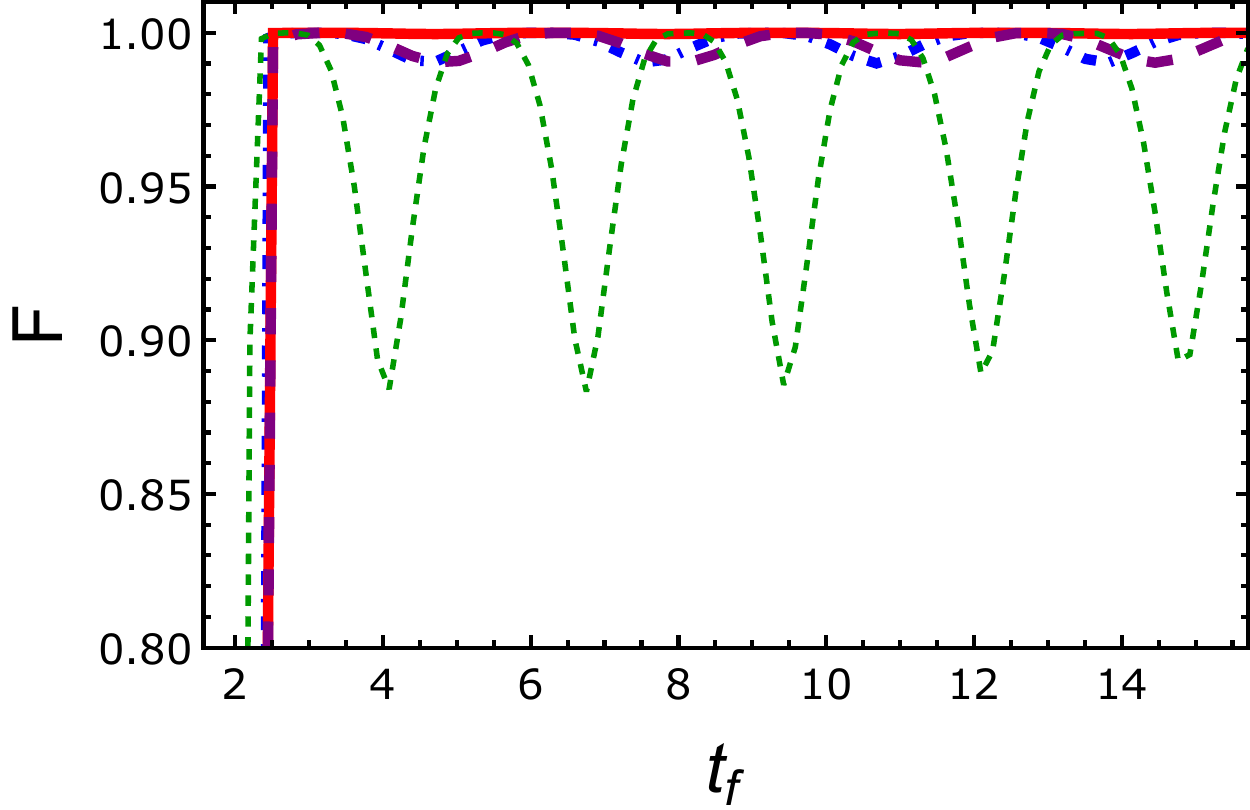}}\\
\caption{Cubic anharmonicity: fidelity with respect to the final time $t_f$ for attractive atomic interactions $g=-0.1$ ( red solid), $g=-0.5$ (dash-dotted blue), {$g=-2$ (dotted green)},
repulsive interaction $g=0.5$ (dashed purple), the anharmonic strength $\kappa=0.02$.}
\label{fig:Fidgtfs}
\end{center}
\end{figure}

To check the performance of the STA trajectories, we define the fidelity at the final time $t_f$ as
\begin{equation}
   F=\vert \langle \psi_{STA}(t_f) \vert \Phi_f \rangle \vert ^2,
\end{equation}
where $\psi_{STA}(t_f)$ is obtained from the direct numerical simulation (split-operator method) of Eq. (\ref{GPE}) using the STA trajectory of $x_0(t)$.
{The desired ground state $\Phi$ is obtained by the imaginary time-evolution technique. $\Phi_{0,f}$ denotes the initial and final ground states, respectively.} 
Noting that we take the ground state $\Phi_0$ as an initial state when we do the time-evolution to get the final state $\psi_{STA}(t_f)$.  
The fidelity of the example in Fig. \ref{figx3}(c) at the final time is $F>0.999$. 
The high performance of fidelity in short time with both attractive and repulsive interactions is plotted in Fig. \ref{fig:Fidgtfs}.
The oscillations are due to the fact that the Gaussian ansatz (\ref{anstz}) is not the solution of BECs with atomic interactions. It is reported that fidelity is improved by using a soliton ansatz in the attractive nonlinear system \cite{Jing17}. 
Thus in this case, the strong attractive interaction will {lead to} the period oscillations (see dotted-green line $g=-2$) due to the Gaussian ansatz we applied in variational approach. 
For the case of repulsive interaction, the period is greater than the attractive one.

\subsection{Quartic anharmonicity}
\label{sec:quartic}

In this section, we shall concentrate on the fast transport of BEC in quartic anharmonicity. The potential reads
\begin{equation}
V(x)= \frac{1}{2} \left(x-x_0\right)^2 +\frac{1}{4!}\lambda(x-x_0)^4.
\end{equation}
Since $\lambda\neq 0$ and $\kappa=0$, the coupled Ermakov-like and Newton-like equations (\ref{engeering1}) become 
\begin{eqnarray}
&&\ddot{a}=\frac{1}{a^3}-a\left(1+18\lambda q^2 \right){+}\frac{gN}{\sqrt{2\pi}a^2}-9\lambda a^3,
\label{a1x4}
\\
&&\ddot{q}=\ddot{x}_0-(1+18\lambda a^2) q-12\lambda q^3.
\label{xc1}
\end{eqnarray}
The first equation (\ref{a1x4}) predicts the breathing mode and the oscillations in the width of the wave packet during the transport.

\begin{figure}[]
{\includegraphics[width=0.35\textwidth,height=4cm]{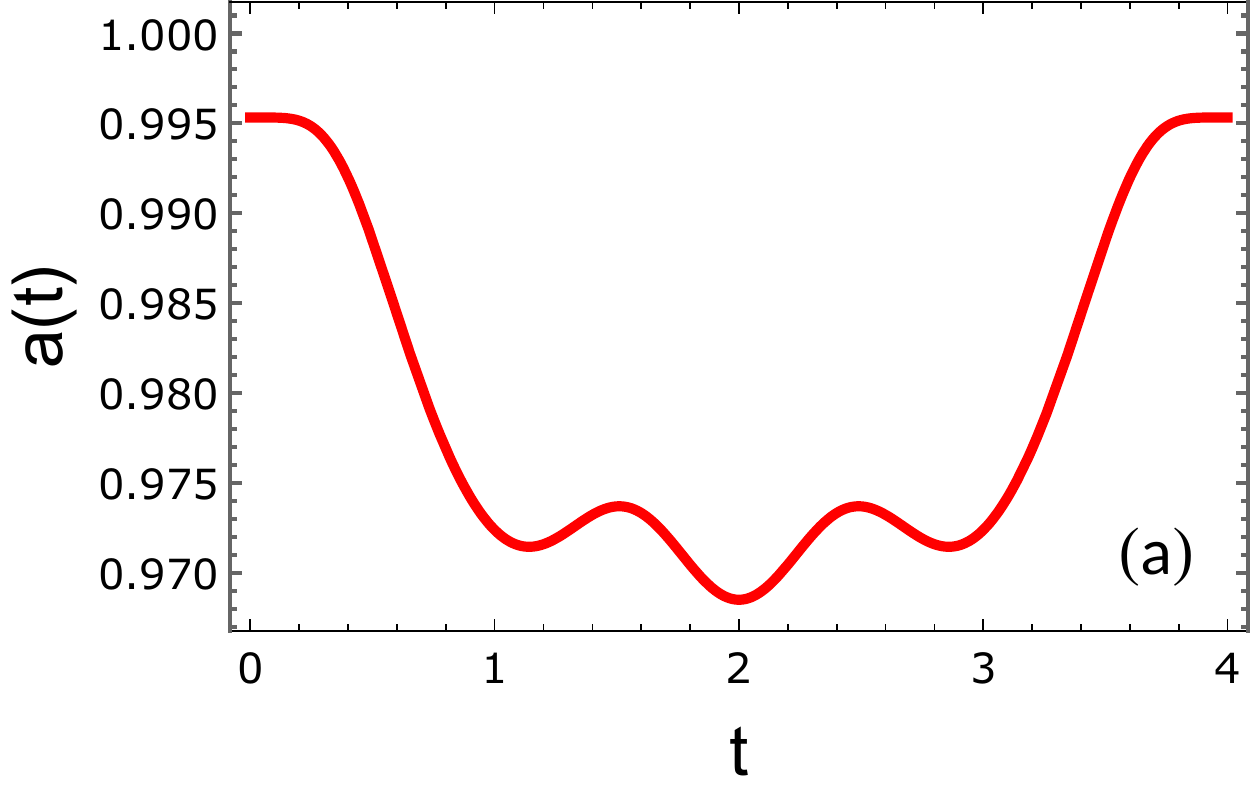}}
{\includegraphics[width=0.35\textwidth,height=4cm]{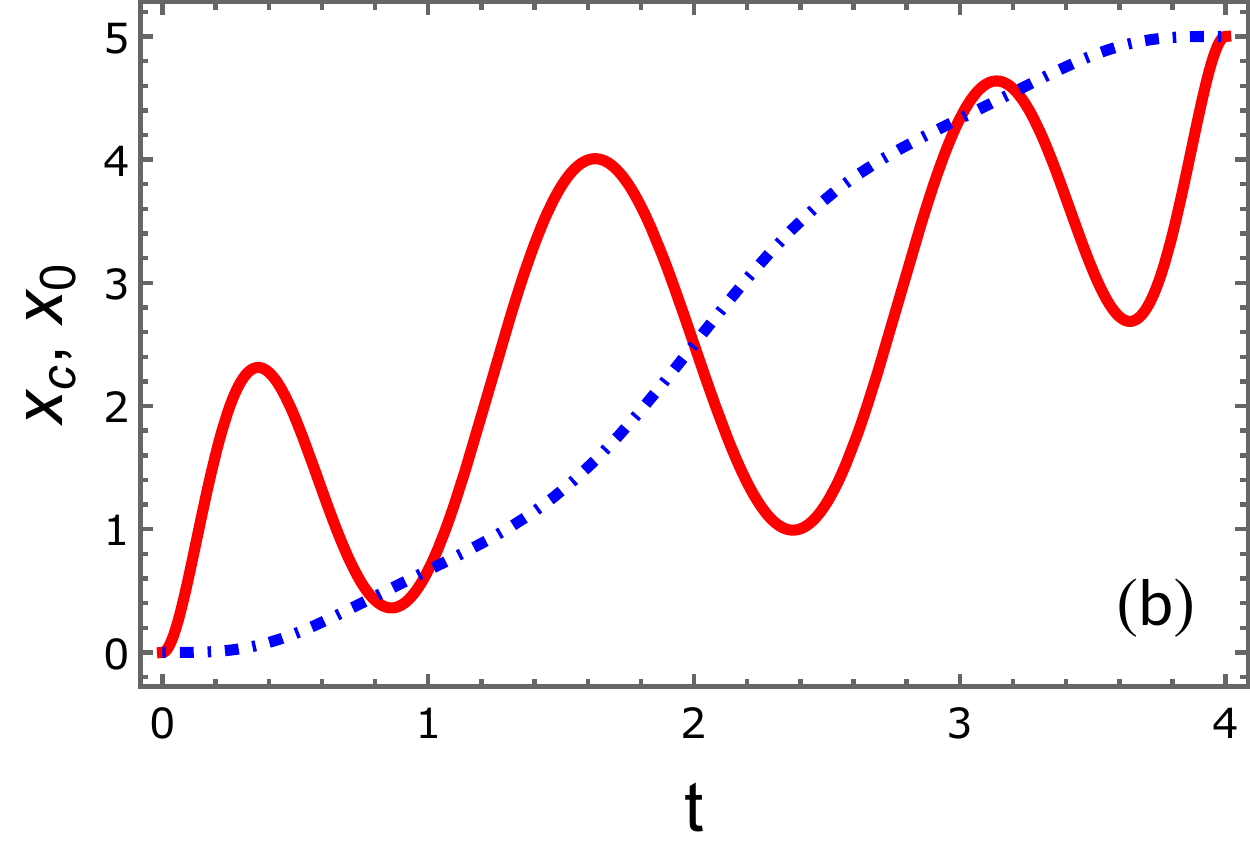}}
\caption{ Quartic anharmonicity: (a) The width $a$ with respect to time.
(b) Shortcut for the designed trajectory for center of mass $x_c$ (dash-dotted blue) and trap center $x_0$ (red solid).
Parameters: $a_{0}=0.995329$, $g=0.01$,  $\lambda=0.06$, $t_f=4$ and $d=5$. }
\label{figx4}
\end{figure}

Our inversion strategy will be different from the one followed previously for cubic anharmonicity because the displacement $q$ appears quadratically in Eq. (\ref{a1x4}). 
We shall design $q(t)$ with a polynomial as $q(t)=\sum_{n=0}^M q_n t^n$.
The boundary conditions for function $q$ in Eq. (\ref{xc1})
\begin{eqnarray}
\label{q0x4}
q(0)=0,\quad q(t_f)=0, 
\nonumber \\
\dot{q}(0)=0,\quad \dot{q}(t_f)=0.
\end{eqnarray}
Then we insert the polynomial function $q$ into the coupled Eqs. (\ref{a1x4}) and (\ref{xc1}) to parameterically solve the functions of width $a$ and $x_c$ with the conditions $a(0)=a_0$, $\dot{a}(0)=0$ and $x_c(0)=0$, $\dot{x}_c(0)=0$. However, we need additional boundary conditions to achieve the final state at final time $t_f$, with
\begin{eqnarray}
a(t_f)=a_0,\quad \dot{a}(t_f)=0, 
\nonumber \\
x_0(t_f)=d,\quad \dot{x_0}(t_f)=0,
\label{x4ax0}
\end{eqnarray}
where $a_{0}$ is the initial and final width calculated by Eq. (\ref{a1x4}) by imposing $\ddot{a}=0$.
The number of the boundary conditions above is eight, therefore one can choose $M=7$. However, we want to demand the following conditions,
\begin{equation}
\label{eta0}
    q\left(\frac{t_f}{4}\right)=0, \quad  q\left(\frac{3t_f}{4}\right)=0,
\end{equation}
to make the distance difference $q$ between the center of potential and the center of wavepacket coincide at these two times.
Alternative boundary conditions would be also possible. According to the above boundary conditions (\ref{q0x4}) - (\ref{eta0}), we obtain the functions $q$, $a$, $x_c$ and $x_0$. An example of the resulting trap trajectory and dynamics is shown in Fig. \ref{figx4}. 
Note that this stationary value makes it different from the transport of cold atoms in purely harmonic traps, since the nonlinearity and anharmonic term
are involved. On the other hand, we shall also emphasize that the width $a$ oscillates (see Fig. \ref{figx4} (a)) during the transport, calculated from Eq. (\ref{a1x4}): {this oscillation is again due to the coupling term between quartic anharmonicity $\lambda$ and width $a$ in the sense that with $\lambda=0$, the solution of Eq. (\ref{a1x4}) will be again a constant width $a$.}
We are now in a position to design the shortcuts to adiabatic transport protocol. Fig. \ref{figx4} (b) shows the trajectories of the center of mass of
wave packet and trap center, by using inverse engineering and boundary conditions, mentioned before. 
At the initial and final times, the trajectories coincide
with each other, which means there is no displacement deviation, guaranteeing the high fidelity ($F=0.9999$) of the transport.

\begin{figure}[]
\begin{center}
\scalebox{0.6}[0.6]{\includegraphics{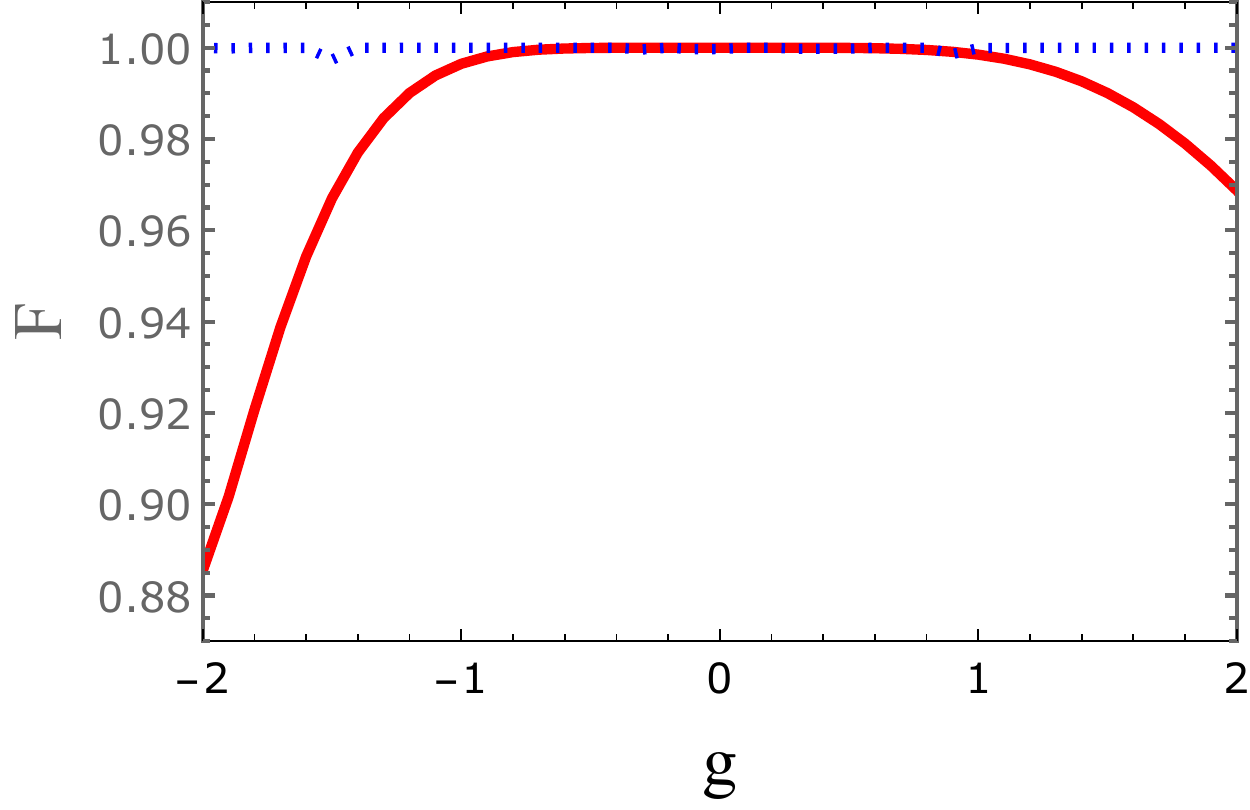}}
\caption{Fidelity for quartic $\lambda=0.06$ (blue-dotted line), and cubic $\kappa=0.02$ (red line) with respect to $g$. Other parameters are $t_f=3\pi$, $N=1$ and $d=5$.}
\label{fig:Fg}
\end{center}
\end{figure}

\section{Effect of nonlinearity }
\label{sec3}

In this section, we shall check the fidelity of our results by solving the GPE numerically 
(without approximation) with the designed shortcuts. 
Fig. \ref{fig:Fg} demonstrates that the fast transport of BECs is perfect with various anharmonic traps taking into account the attractive and repulsive interactions. 
{The Gaussian ansatz is also valid for the variational approximation in our model in the presence
of an atomic interaction $g \neq 0$. In Fig. \ref{fig:Fg}, the fidelity for cubic anharmonicity $\kappa = 0.02$ is plotted for different atomic interactions $g$. 
The fidelity is above $0.99$ for atomic interaction $|g|<1.2$, i.e. fast transport of BEC in cubic anharmonic traps can be achieved for both attractive and repulsive interactions.
The fidelity drops with interactions $|g| \ge 1.2$.
This is not surprising as
one would expect that the Guassian variational approach (\ref{anstz}) works better for small interaction $g$.
For example, the nonlinearity $g =-2$ is in the range where we would not expect the ansatz to work.
For $g=-2$, the fidelity will oscillate with respect to the final time which is shown in Fig. \ref{fig:Fidgtfs}. 
In Fig. \ref{fig:Fg}, the fidelity for quartic anharmonicity $\lambda = 0.06$ is plotted for different atomic interactions $g$. 
The fidelity is always greater than $0.99$, i.e. fast transport of BEC in quartic anharmonic traps can be achieved for both attractive and repulsive interactions.
 }

\section{Conclusion}

In summary, we present an efficient way to design high-fidelity and fast transport of BEC in anharmonic traps by combining the variational approach and inverse engineering methods.
The shortcuts to adiabatic transport of the BEC are demonstrated with numerical examples in quartic and cubic anharmomonicity traps.
It is concluded that perfect transport can be achieved in cubic anharmonic traps in the presence of both attractive and repulsive interactions.
Our method presented here is different from the previous ones \cite{Qi2015}, in which the anharmonic potential is considered as a perturbation. The shortcut trajectory can be further optimized by using optimal control theory, for instance, by taking into account noise and error in traps position and frequency \cite{Lu2018}.
The technique may be extended to 3D Gaussian-beam optical traps\cite{Andreas123D}, the spin-orbit coupled BECs\cite{jing22}, strongly interacting bosons (Tonks-Girardeau gas) \cite{Haque}, and superfluid Fermi gas \cite{FermiGas}.
The transport of soliton matter waves will also be reported in future work.
We expect our shortcut design for fast transport to have potential applications not only in atom interferometry \cite{Nivet} but also in quantum information processing.

\section*{Acknowledgment}

We are grateful to D. Rea, C. Whitty and M. Odelli for commenting on the manuscript. 
J.L. appreciated the discussions from J. G. Muga and D. Gu\'ery-Odelin  at early stage of work.
J.L. and A.R. acknowledge that this publication has emanated from
research supported in part by a Grant from Science Foundation Ireland
under Grant number 19/FFP/6951 (``Shortcut-Enhanced Quantum
Thermodynamics"). 
This work has been financially supported by EU FET Open Grant  EPIQUS (899368),  QUANTEK project (KK-2021/00070),  the Basque Government through Grant No. IT1470-22, and the project grant 
PID2021-126273NB-I00 funded by MCIN/AEI/10.13039/501100011033 and by
``ERDF A way of making Europe" and ``ERDF Invest in
your Future". X.C. acknowledges the Ram\'on y Cajal program (RYC-2017-22482).

%



\end{document}